\documentclass[Physsubmission, Phys]{SciPost}
\hypersetup{
    colorlinks,
    linkcolor={red!50!black},
    citecolor={blue!50!black},
    urlcolor={blue!80!black}
}
\usepackage{soul}
\usepackage[makeroom]{cancel}
\usepackage[bitstream-charter]{mathdesign}
\DeclareSymbolFont{usualmathcal}{OMS}{cmsy}{m}{n}
\DeclareSymbolFontAlphabet{\mathcal}{usualmathcal}

\newcommand{\beqa}{\begin{eqnarray}}
\newcommand{\eeqa}{\end{eqnarray}}
\newcommand{\eK}{e^{\big|\langle z \rangle \big|^2}}
\newcommand{\eKs}{e^{  \frac12 \big|\langle z \rangle \big|^2}}
\newcommand{\pp}[1]{(\langle \varphi^{#1} \rangle + \varphi^{#1})}

\newcommand{\pb}[1]{(\langle \varphi^\dag_{#1} \rangle + \varphi^\dag_{#1})}
\newcommand{\bS}[1]{(\langle S^{#1} \rangle + S^{#1})}
\newcommand{\bSb}[1]{(\langle S^\dag_{#1} \rangle + S^\dag_{#1})}

\newcommand{\lag}{\langle}
\newcommand{\rag}{\rangle}
\newcommand{\lb}{\big|}

\newcommand{\nn}{\nonumber}

\begin{document}

\begin{center}{\Large \textbf{
New perspectives in Gravity-Mediated Supersymmetry Breaking\\
}}\end{center}

% TODO: write the author list here. Use initials + surname format.
% Separate subsequent authors by a comma, omit comma at the end of the list.
% Mark the corresponding author with a superscript *.
\begin{center}
Robin Ducrocq\textsuperscript{1}*
\end{center}

\begin{center}
{\bf 1} Theory Group, IPHC, Strabourg, France
\\
* robin.ducrocq@iphc.cnrs.fr
\end{center}

\begin{center}
\today
\end{center}

% For convenience during refereeing (optional),
% you can turn on line numbers by uncommenting the next line:
%\linenumbers

\definecolor{palegray}{gray}{0.95}
\begin{center}
\colorbox{palegray}{
  \begin{tabular}{rr}
  \begin{minipage}{0.1\textwidth}
    \includegraphics[width=20mm]{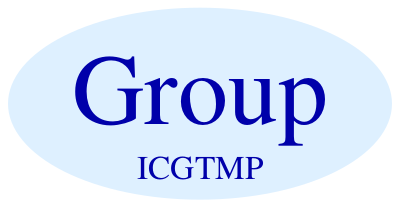}
  \end{minipage}
  &
  \begin{minipage}{0.85\textwidth}
    \begin{center}
    {\it 34th International Colloquium on Group Theoretical Methods in Physics}\\
    {\it Strasbourg, 18-22 July 2022} \\
    \doi{10.21468/SciPostPhysProc.?}\\
    \end{center}
  \end{minipage}
\end{tabular}
}
\end{center}

\section*{Abstract}
{\bf
% TODO: write your abstract here.
New solutions in supersymmetry breaking through gravity mediation have been recently discovered. Such solutions have interesting properties regarding renormalisation and introduce new contributions in the scalar potential that may help to resolve some issues of the Standard Model. The purpose of this article is to investigate the consequences of these new structures. We construct a model related to these new solutions, the S2MSSM, and present some preliminary results on the effects of these new contributions, especially on the Standard Model's Higgs boson mass.}
%Supersymmetry must be broken in order to be coherent with observations in particle physics experiments.

%\vspace{10pt}
%\noindent\rule{\textwidth}{1pt}
%\tableofcontents\thispagestyle{fancy}
%\noindent\rule{\textwidth}{1pt}
%\vspace{10pt}

\section{Introduction}
\label{sec:intro}

The Standard Model of particle physics provides a robust framework to describe the behaviour of particles and fundamental interactions. However, several issues still remain in this model. Some of these problems may be solved by embedding the Standard Model in a more fundamental theory. There exist several possibilities. Two of them are supersymmetry and its local version, supergravity. Such theories are defined within the framework of Lie superalgebras in the line of the \sloppy Haag–Lopuszański–Sohnius theorem \cite{HLS}. This theorem strongly constrains the possible symmetries of the spacetime. In the simplest case ($N=1$ supersymmetry and supergravity), this theorem restricts the symmetry to be:
\begin{equation}
\mathfrak{g} = g_0\oplus g_1 \quad \text{with}\quad  g_0 = \mathfrak{Iso}(1,3)\times \mathfrak{g}_C\quad ,\quad g_1=S_L\oplus S_R \nn
\end{equation}
with $\mathfrak{Iso}(1,3)$ the Poincaré algebra, $\mathfrak{g}_C$ a compact Lie algebra related to internal symmetries and $S_L$ (resp. $S_R$), the left-(right-)handed spinor representation where $S_L=\{Q_\alpha,\ \alpha=1,2\}$ ($S_R=\{\bar{Q}^{\dot{\alpha}},\ \dot{\alpha}=1,2\}$) and $(Q_\alpha)^\dag = \bar{Q}_{\dot{\alpha}}$. The subspace $\mathfrak{g}_0$ is called even whereas $\mathfrak{g}_1$ is called odd.\medskip

However, the spectrum of such theory is incompatible with the actual measurements. Supersymmetry and supergravity must then be broken. Several consistent mechanisms exist in supergravity. We focus on one of them, namely, gravity-mediated supersymmetry breaking. In such scenarios, supergravity is assumed to be broken in a hidden sector. Such a configuration induces, through gravitational effects, supersymmetry breaking in the usual field sector. Solutions to this mechanism were first classified in the 80s \cite{soni_analysis_1983}. Recently, new solutions have been discovered \cite{moultaka_low_2018} with new supersymmetry breaking terms and a new field sector with specific properties. \medskip

After a general presentation of the gravity-mediated supersymmetry breaking mechanism, we present the new solutions. A particular model, the S2MSSM, is then constructed. The mass matrix of the scalar sector is finally analysed.

\section{Supersymmetry Breaking through gravitational interactions}
To construct a model in supergravity, we choose a gauge group $G$. In our case, we consider $G=SU(3)_c\times SU(2)_L\times U(1)_Y$. Vector superfields associated to the strong $SU(3)_c$ and the electroweak $SU(2)_L\times U(1)_Y$ interactions are naturally introduced in the adjoint representation of $G$. We also include a matter sector with chiral superfields subdivided into two subsectors. The first is the visible (or observable) sector $\{\Phi^a,\ a=1,\dots,n_a\}$ where $\Phi^a=(\phi^a,\chi_\phi^a,F_\phi^a)$,\footnote{We denote the components of a chiral superfield $X^A$ by $X^A=(x^A,\chi_X^A,F_X^A)$ with $x^A$ a scalar field, $\chi_X^A$ a left-handed Weyl spinor and $F_X^A$ an auxiliary field.} containing the usual fields of the Standard Model with their associated supersymmetric partners. The second is the hidden sector $\{Z^i,\ i=1,\dots,n_i\}$ with $Z^i=(\zeta^i,\chi_\zeta^i,F_\zeta^i)$. Finally, two gauge invariant functions of the chiral superfields are introduced: a real function called the Kähler potential $K$ which leads to the kinetic term of chiral superfields and a holomorphic function, the superpotential $W$, which generates the Yukawa couplings of the Standard Model. Supergravity is then assumed to be broken in the hidden sector with $\langle \zeta^i \rangle = \mathcal{O}(m_p)$ where $m_p$ is the Planck mass (we also have $\langle \phi^a \rangle \ll m_p$). The $F$-term of the scalar potential of supergravity can then be computed:
\begin{equation}
V_F = \exp\Big(K/m_p^2\Big)\Big(\mathcal{D}_AW(K^{-1})^{A}{}_{B^\ast}\mathcal{D}^{B^\ast}\bar{W} - \frac{3}{m_p^2}|W|^2 \Big) \label{eq:Vsugra}
\end{equation}
with:
\begin{equation}
\mathcal{D}_AW = \partial_AW +\frac{W}{m_p^2}\partial_AK \ , \quad (K^{-1})^{A}{}_{B^\ast} = \Bigg( \frac{\partial^2 K}{\partial X^A\partial X^\dag_{B^\ast}}\Bigg)^{-1} \nn
\end{equation}
($\{X^A\}=\{Z^i,\Phi^a\}$). Considering the low energy limit (\textit{i.e.}, taking $m_p\rightarrow \infty$), we obtain the classical potential of supersymmetry $V_{SUSY}$ with additional terms $V_{\cancel{SUSY}}$ which explicitly break supersymmetry:
\begin{equation}
V_F = V_{SUSY} + V_{\cancel{SUSY}} \ .\label{eq:Vbreak}
\end{equation}
Note that the form of the superpotential and the Kähler potential must not induce dangerous couplings in Eq.~\ref{eq:Vsugra}. Indeed, at low energy, couplings proportional to the Planck mass $m_p$ generate instabilities in the matter sector. We must therefore impose that the interactions in the visible sector must be proportional in the potential to $m_p^n$ with $n\leq 0$.\medskip

We are interested in solutions for which the Kähler potential K and the superpotential W can be expanded as power of the Planck mass:
\begin{equation}
K(Z,Z^\dag,\Phi,\Phi^\dag)=\sum\limits_{n=0}^{r}K_n(Z,Z^\dag,\Phi,\Phi^\dag)m_p^n \ , \quad W(Z,\Phi)=\sum\limits_{n=0}^{s}W_n(Z,\Phi)m_p^n \label{eq:power_mp}
\end{equation}
(we thus exclude no-scale solutions). Under these assumptions, one obtain two solutions using a canonical Kähler potential. The first one is the historical solution discovered by Soni \& Weldon \cite{soni_analysis_1983}, which is the cornerstone for all the studies involving gravity-mediated supersymmetry breaking up to now. The second is a new structure that will be described in the next section.

\section{New solutions in Gravity-Mediated Supersymmetry Breaking}
We briefly present the new solutions \cite{moultaka_low_2018} associated to a canonical Kähler potential:
\begin{equation}
K(Z,Z^{\dagger},\Phi,\Phi^{\dagger}) = Z^iZ^{\dagger}_i + \Phi^a\Phi^{\dagger}_a \ . \nn
\end{equation}
Following Eq.~\ref{eq:power_mp}, two possible forms for the superpotential have been identified. The first corresponds to the known solution developed in \cite{soni_analysis_1983}. The second has a new structure and introduces a new singlet superfield sector $\{\mathcal{S}^p,\ p=1,\dots,n_p\}$ (with ${\cal S}^p=(S^p,\chi_S^p,F_S^p)$):
\begin{equation}
W(Z,\Phi,\mathcal{S}) =  m_p W_1(Z,\mathcal{S}) + W_0(Z,\Phi,\mathcal{S})\label{eq:Wmp}
\end{equation}
with:
\begin{equation}
W_1(Z,\mathcal{S}) = W_{1,0}(Z) + W_{1,p}(Z)\mu^{*}_p\mathcal{S}^p\ , \quad W_{0}(Z,\Phi)=W_{0,p}(Z)\mathcal{S}^p + W_0 (Z, \mathcal{U},\Phi ) \label{eq:W}
\end{equation}
and
\begin{equation}
{\cal U}^{pq} = \mu^p \mathcal{S}^q - \mu^q\mathcal{S}^p \ .\label{eq:U}
\end{equation}
The functions $W_{1,0}$, $W_{1,p}$, $W_{0,p}$ and $W_{0}$ are holomorphic functions of chiral superfields. The form of $W(Z,\Phi,\mathcal{S})$ in Eqs.~\ref{eq:Wmp} \& \ref{eq:W} and $\mathcal{U}$ in Eq.~\ref{eq:U} is dictated to avoids dangerous couplings between the visible and the hidden sector in the low energy limit. The new singlet superfield sector $\{\mathcal{S}^p\}$ is called ‘‘hybrid’’. It involves in $W_1(Z,\mathcal{S})$ a term proportional to the Planck mass $m_p$ (see Eq.~\ref{eq:Wmp}), but still produces a divergent-free low energy potential. The low energy scalar potential Eq.~\ref{eq:Vbreak} associated with Eqs.~\ref{eq:Wmp} and \ref{eq:W} is:
\begin{equation}
V = V_{SUSY} + \Lambda m_p^2 + V_{SOFT} + V_{HARD} \label{eq:Vbreak}
\end{equation}
where $\Lambda$ is the cosmological constant. The two remaining terms break supersymmetry explicitly. The terms $V_{SOFT}$ are \textit{soft} supersymmetric breaking terms, \textit{i.e.}, terms that lead to logarithmic divergences through loop corrections. Such contributions are already present in the historical classification. The general form of $V_{SOFT}$ has been classified \cite{GIRARDELLO198265} and takes the form:
\begin{equation}
V_{SOFT} = \Big( C_i\tilde{\phi}^i + \frac12 B_{ij}\tilde{\phi}^i\tilde{\phi}^j + \frac16 A_{ijk}\tilde{\phi}^i\tilde{\phi}^j\tilde{\phi}^k + \text{h.c.} \Big) + m_{\tilde{\phi}}^2 \tilde{\phi}^i\tilde{\phi}^{\dagger}_i \label{eq:vsoft}
\end{equation}
with $\{\tilde{\phi}^i\}=\{\phi^a,S^p\}$. The first terms are holomorphic while the last term is real and correspond to the mass term to each chiral fields $\phi^i$. Note that the parameters $C_i,\ B_{ij}$ and $A_{ijk}$ are not arbitrary but are related to the form of the superpotential $W_0$ in Eq.~\ref{eq:W} (which is polynomial of degree three). \medskip

The specific structure of the S-sector generates the last term in Eq.~\ref{eq:Vbreak}. Such couplings are \textit{hard} breaking terms, \textit{i.e.}, induce quadratic loop divergences. The general form of the \textit{hard} breaking terms takes the form:
\begin{align}
V_{HARD} =&\  \Big( \big( D_{i}^p\phi^i + \frac12 E_{ij}^p\phi^i\phi^j + \frac16 F_{ijk}^p\phi^i\phi^j\phi^k\big) S^\dag_p + G_{ijk}{}^{l}\phi^i\phi^j\phi^k\phi^{\dagger}_l+ H_{ijp}{}^{l}\phi^i\phi^jS^p\phi^{\dagger}_l + \text{h.c.} \Big) \nn\\
&+ Q_{i,p}{}^q \phi^i\phi^\dag_i S^p S^\dag_q + T_{i,p}{}^q S^p S^\dag_p S^r S^\dag_q \ . \nn
\end{align}
The presence of \textit{hard} breaking terms in the potential is new. Such terms differ from \textit{soft} breaking terms since couplings between holomorphic and anti-holomorphic superfields are present. These couplings allow to close S-loops and induce new contributions to the mass of fields $\phi$. Since such hard terms are suppressed by an intermediate scale, the quadratic divergences are reduced and may be sizeable to solve some actual issues of the Standard Model. The \textit{hard} parameters $D^p_i$, $E^{p}_{ij}$ and $F^p_{ijk}$ are also correlated with the holomorphic \textit{soft} breaking terms through the hybrid fields couplings. 

\section{Hybrid extension of the MSSM: the S2MSSM}
In the previous section, we have presented new solutions obtained from a canonical Kähler potential. It is desirable to extend the analysis to the non-canonical case to get a richer mass spectrum. Thus, along the lines of the results of Brignole, Ibanez \& Munoz \cite{BRIGNOLE} and Guidicce \& Masiero \cite{GIUDICEMASIERO}, we have considered a solution assuming a non-canonical Kähler metric. This enables us to identify a possible extension of the Minimal Supersymmetric Standard Model (MSSM) \cite{FAYET1976159} involving hybrid fields $S^p$.
\subsection{Definition of the model}

We assume a hidden sector containing one superfield $Z=(\zeta,\chi_\zeta,F_\zeta)$ and the observable sector of the MSSM $\{\Phi^a\}=\{\Phi^a\}_{MSSM}$. This model is the simplest supersymmetric extension of the Standard Model. The visible sector contains superfields associated to quarks, leptons and the two $SU(2)$ superfields Higgs doublets $H_U$ and $H_D$. We also introduce a hybrid sector $\{\mathcal{S}^p\}$ ($p=1,\dots ,n_p$). Following the results above, the superfield $\mathcal{U}$ is the only superfield that couples to the observable sector. Among these $n_p$ hybrid superfields, we assume that only two superfields $\mathcal{S}^1$ and $\mathcal{S}^2$ interact with $\{\Phi^a\}$ via ${\cal U}$:
\begin{equation}
{\cal U} = \mu^1\mathcal{S}^2 - \mu^2\mathcal{S}^1\ . \nn
\end{equation}
The $n_p-2$ other fields $\{S^3,\dots, S^{n_p}\}$ will play an important role as we will see later. The superpotential and the Kähler potential are:
\begin{align}
&W(\Phi,\mathcal{S},Z) = m_p\big( W_{1,0}(Z) + \mathcal{S}^p\mu_p^\ast W_{1,p}(Z)\big) + \mathcal{S}^pW_{0,p}(Z) + W_{0}(\Phi,{\cal U},Z) \nn \\
&K(\Phi,\Phi^\dag,\mathcal{S},\mathcal{S}^\dag,Z,Z^\dag) = m_p^2 \hat{K}(Z,Z^\dag) + \mathcal{S}^\dag_pS^p + \sum\limits_{a} \Lambda_a(Z,Z^\dag)\Phi^\dag_a\Phi^a \nn
\end{align}
where: 
\begin{equation}
W_0(\Phi,{\cal U},Z) = \lambda(Z){\cal U}H_U\cdot H_D + \frac16 \kappa(Z) {\cal U}^3 + W_{MSSM} |_{\mu = 0} \nn
\end{equation}
with $W_{MSSM}|_{\mu = 0}$, the superpotential of the MSSM (not given here) where the quadratic Higgs doublets coupling is not present. Such a superpotential contains then only Yukawa couplings, \textit{i.e.}, cubic terms. The matrix $\Lambda_a(Z,Z^{\dagger})$ leads to a non-universality of the breaking terms in the usual matter sector. Since the hybrid superfields $\mathcal{S}^p$ are gauge invariant, quadratic and linear contributions can be added. However, we restrict ourselves to a $\mathbb{Z}_3$-invariant $W_0(\Phi,{\cal U},Z)$ superpotential (only cubic couplings) assuming superconformal invariance. \medskip

As seen previously, such solutions generate \textit{soft} and \textit{hard} terms that affect the mass spectrum of particles at the tree level and through loop corrections. We now investigate the mass matrix of such a model.

\subsection{A simple case}
This theory contains many fields. It is then difficult to determine the set of parameters leading to interesting results. In order to find the optimal configuration, we first analyse a simplified model. We assume that only the scalar field from the hidden sector gets a nonzero vacuum expectation value (or \textit{v.e.v.}) with $\langle \zeta \rangle = \mathcal{O}(m_p)$.\footnote{Note that fields from the observable sector can develop a nonzero \textit{v.e.vs} but much smaller than the Planck mass, \textit{i.e.}, $\langle S^p\rangle \ll \langle \zeta \rangle$ and  $\langle \phi^a\rangle =M_{\phi}\ll \langle \zeta \rangle$ (with $M_\phi = M_{EW}\approx 10^2\ \text{GeV}$ or $M_{GUT}\approx 10^{16}\ \text{GeV}$). The effect of these nonzero \textit{v.e.vs.} are taken in account in Section \ref{sec:towardS2}.} We also assume:
\begin{equation}
W_{0}(\Phi,{\cal U},Z) = W_{0}(\Phi,Z)\ . \nn
\end{equation}
The S-sector then only contributes through the two components $W_{1,p}$ and $W_{0,p}$ of the superpotential Eq.~\ref{eq:W}. \medskip

We compute the scalar potential Eq.~\ref{eq:Vsugra} following these hypotheses. Such a potential can be written as Eq.~\ref{eq:Vbreak}. The vanishing of the cosmological constant and the minimisation of the potential are also imposed:
\begin{equation}
\langle V \rangle = 0 \ , \quad \Bigg\langle \frac{\partial V}{\partial X^A} \Bigg\rangle = 0  \ , \quad \Bigg\langle \frac{\partial V}{\partial X^\dag_{A^\ast}} \Bigg\rangle = 0 \label{eq:constraints}
\end{equation}
(with $X^A=\{z,\phi^a,S^p\}$, the scalar part of the chiral superfields and we introduce $z=\zeta/m_p$). These relations highly constrain the parameter space. \medskip

The mass matrix of this model is a $(n_a+n_p+1)\times(n_a+n_p+1)$ matrix mixing the three different sectors. It can be shown that it decouples into two submatrices related to the two sectors $\{\Phi^a\}$ and $\{S^p,z\}$ at first order of $1/m_p^2$, and so can be diagonalised separately. The mass matrix $\mathbb{M}'^2$ in the sector $X'^A=\{S^p,z\}$ reads:
\beqa
\mathbb M'^2 = \Bigg\langle \frac{\partial^2 V}{\partial X'^AX'^\dag_{B^\ast}}\Bigg\rangle =\begin{pmatrix} \delta_p{}^q m_{3/2} + b{\cal I}_p\bar{\cal I}^q & c{\cal I}_p\\
  \bar{c}\bar{\cal I}^q & d \\
  \end{pmatrix}\quad \text{with}\quad m_{3/2}=\frac{1}{m_p^2}e^{\langle K\rangle/m_p^2} \langle W \rangle 
  \label{eq:matrixprime}
\eeqa
where ${\cal I}_p = \big\langle \mu^\ast_p W_{1,p}(z)m_p + W_{0,p}(z) \big \rangle$, b,c and d are some constants related to the parameters of the superpotential, and $m_{3/2}$ is the gravitino mass. Performing a change of basis, we rewrite the mass matrix $\mathbb{M}'^2$ in the form:
\beqa
\mathbb M'^2 = \begin{pmatrix} m_{3/2} \mathbb{I}_{n-1}&0&0\\
  0&m_{3/2} + b |{\cal I}|^2&c |{\cal I}|\\
  0 &\bar c |{\cal I}|&d\\
  \end{pmatrix}\quad \text{with}\quad \sum\limits_{p=1}^{n_p}{\cal I}_p\bar{\cal I}^p=|{\cal I}|^2 \ .
  \label{eq:matrixprime_change}
\eeqa
We then obtain one S-field mixing in a non-trivial way with the hidden field $z$ and the $n_p-1$ remaining S fields with a mass equal to the gravitino mass $m_{3/2}$. Proving inductively that $\text{Tr}[(\mathbb{M}'^2)^n]$ (with $n \in \mathbb{N}$) is only a function of $|{\cal I}|^2$, one can show that the eigenvalues do not depend on ${\cal I}_p$ thanks to the vanishing of the cosmological constant:
\begin{equation}
\label{eq:Lambda}
\left \langle V \right \rangle = e^{|\langle z \rangle|^2}\Big( |{\cal I}|^2 + \sum \limits_a \big|\langle \partial_a W_0\rangle\big|^2\Big)+ m_p^2
( |m'_{3/2}|^2 -3|m_{3/2}|^2) = 0 \ \text{with} \ m'_{3/2} = \frac{1}{m_p^2}e^{\langle K\rangle/m_p^2} \langle \mathfrak{d}_z W \rangle  \nn
\end{equation}
where $\mathfrak{d}_z W = \partial_zW + z^{\dag}W$. \medskip

To understand the effects of the hybrid sector $\{{\cal S}^p\}$ on the Standard Model, we have investigated the consequences of this new sector on the Higgs boson mass. Several points can be mentioned:
%This new sector's impact on the visible sector's mass spectrum has also been examined. In this simple case, there is no direct coupling between these two field sectors in the superpotential $W_0$. However, loop contributions modify the tree-level masses of the visible sector. In this sense, we have scrutinised the order of magnitude of the one-loop contribution on the Higgs boson. Two points have been highlighted. In order to reduce the fine-tuning of the Higgs boson mass, the visible sector must be embedded in a GUT model to increase the visible sector's energy scale along with the loop contribution. The second point is that the hidden-sector must be highly fine-tuned and contains several fields to increase the Higgs boson mass to $125\ \text{GeV}$. For more information on this analysis, see \cite{phdthesis}.
\begin{itemize}
\item Since there are no interactions between the hybrid and the observable sector in $W_0$, the only contributions of the S-sector on the Higgs boson mass are obtained through the \textit{hard} breaking terms and thus through loop-corrections.
\item The order of magnitude of the one loop-correction is proportional to the energy scale of the visible sector. In order to increase such contributions, we embed the Standard Model in a GUT model such that $\langle \phi^a\rangle \approx M_{GUT}$.
\item A quantitative study on the order of magnitude of the one S-loop contribution to the Higgs boson mass has been done. With such hypotheses, we have highlighted several configurations leading to a Higgs boson mass of $125\ \text{GeV}$. Nevertheless, such configurations require a certain level of fine-tuning in the hidden sector. 
\end{itemize}
This study enables us to put in evidence the correct strategy for analysing the S2MSSM without the simplifying assumptions imposed previously.

\subsection{Towards the general S2MSSM}\label{sec:towardS2}
We now reintroduce the dependence of ${\cal U}$ in the superpotential $W_{0}(\Phi,{\cal U},Z)$ Eq.~\ref{eq:W}. Non-null vacuum expectation values for the hybrid fields $\langle S^p\rangle \neq 0$ and the visible sector $\langle \phi^a\rangle \neq 0$ are also assumed. The energy scale of the $\Phi$-sector corresponds to the electroweak scale ($M_{EW}\approx 10^2\ \text{GeV}$) or a GUT scale ($M_{GUT}\approx 10^{16}\ \text{GeV}$).\medskip

The complete form of the scalar potential of the S2MSSM is given in Appendix \ref{sec:VS2MSSM}. The mass matrix in the sub-sector $\{S^p,z\}$ can be written in the following form:
\beqa
\mathbb M'^2 = \begin{pmatrix} \delta_p{}^q a' + e' + b'{\cal J}_p\bar{\cal J}^q & c'{\cal J}_p + f'_p\\
  \bar{c'}\bar{\cal J}^q + \bar{f'}^q & d' \\
  \end{pmatrix} \quad \text{with} \quad {\cal J}_p = {\cal I}_p + \langle \partial_pW_0\rangle 
  \label{eq:matrixprime_general}
\eeqa

with $a'$, $b'$, $c'$, $d'$, $e'$ and $f'_p$ some constants. Note that $e',\ f'$ and $d'$ depend on the parameter ${\cal J}_p$. Due to these new contributions, the simple structure $b{\cal I}_p\bar{\cal I}^q$ in the $\{S^p\}$-sector (see Eq.~\ref{eq:matrixprime}) is lost. Consequently, the spectrum is not degenerate with a mass equal to $m_{3/2}$. \medskip

Mention again that a complete qualitative analysis is tedious due to the number of new contributions in the scalar potential (see Appendix \ref{sec:VS2MSSM}). A numerical computation of the mass matrix is necessary to find configurations that reduce the mass matrix to a form equivalent to Eq.~\ref{eq:matrixprime_change}. Such a study is in progress \cite{lowsugra2}. Such new terms may also help to resolve two actual issues in the Standard Model and in supersymmetry, \textit{i.e.}:
\begin{itemize}
\item reduce the fine-tuning on the Higgs boson mass through tree-level and loop corrections and help to naturally obtain a mass near $125\ \text{GeV}$,
\item push the squark masses to higher energy which may explain the non-detection of supersymmetry in particle physics experiments.
\end{itemize}
Note also that this model can have an interesting relationship with a model called NMSSM \cite{Ellwanger_2010} (extension of the MSSM with one singlet superfield). The relation between these two models is also under investigation \cite{lowsugra2}.
\section{Conclusion}
New solutions where supersymmetry is broken through gravitational mediation involving \textit{hard} breaking terms have been investigated. The contributions of hard breaking terms have been studied in this paper through the construction of a model related to these new solutions, the S2MSSM. The mass spectrum of this new model has been calculated assuming the vanishing of the cosmological constant, \textit{i.e.}, $\langle V \rangle =0$. Following some simplifying assumptions, we obtain in the particle spectrum several degenerated states with a mass equal to the gravitino mass. However, such structure is lost when assuming all the contributions in the S2MSSM.\medskip

A complete numerical analysis through a spectrum generator may be useful to investigate all the (tree-level and loop-level) contributions of the new field sector $\{\mathcal{S}^p\}$.

%You must include a conclusion.
\section*{Acknowledgements}
I'm grateful to the organisers of the 34th International Colloquium on Group Theoretical Methods in Physics to allowed me to present these results in a parallel session.
%Acknowledgements should follow immediately after the conclusion. 

\begin{appendix}
\section{Scalar potential of the S2MSSM} \label{sec:VS2MSSM}
The purpose of this appendix is to give the scalar potential of the S2MSSM in the low energy limit. We define $W_{0}=M_4^3\omega_0$ and $\phi^a=M_4\varphi^a$ with $M_4=M_{EW}$ or $M_{GUT}$. We also introduce the notation 
\begin{equation}
\Delta f(z,S,\Phi) = f(\langle z\rangle ,S + \langle S\rangle ,\Phi + \langle \Phi\rangle ) - f(\langle z\rangle ,\langle S\rangle ,\langle \Phi\rangle )\ .\nn
\end{equation}
The definition of ${\cal I}_p$ is given in Eq.~\ref{eq:matrixprime}. The complete form of the scalar potential is:
{\small 
\begin{align} 
\label{eq:geneV}
V&= m_p^2\lb m_{3/2}\lb^2 \big(\frac 1{\lb \xi_{3/2}\lb^2} -3 \big)+ \eK\Big(\sum \limits_p
\lb {\cal I}_p +M_4 ^3\partial_p \omega_0 \lb^2 + M_4^4 \partial_a \omega_0 \partial^{a^\ast}\bar \omega_0 \lag (\Lambda^{-1})^a{}_{a^\ast}\rag\Big)\nn\\
 & + \Big(\bS{p} \bSb{p}\Big)\Big( \lb m_{3/2}\lb^2 T+ 
 \frac 1{m_p^2} \eKs \big[\bar m_{3/2}S^r T_r + \mbox{h.c.} \big]+\frac 1{m_p^4} \eK S^r S^\dag_t T^t{}_s\Big)  \nn\\
 &+\frac 1{m_p^2} \eK S^p S^\dag_q \Big( \mathfrak{d}_z {\cal I}_p 
\mathfrak{d}^z  \bar{\cal I}^q -3 {\cal I}_p \bar{\cal I}^q\Big) + \eKs \bar m_{3/2}S^p \big(\frac1 {\bar \xi_{3/2}} \mathfrak{d}_z {\cal I}_p -3 \ {\cal I}_p + \mbox{h.c.}\big) \nn\\ 
& +\frac 1{m_p^2} \eKs \Bigg\{ \Big(M_4^2 \pb{a^\ast} \pp{a} \lag \Lambda^{a^\ast}{}_a \rag +\bSb{p} \bS{p}\Big) \bS{q} {\cal I}_q \times\nn\\
& \hskip 0.5truecm  \Big(\bar m_{3/2}+ \frac 1 {m_p^2} \eKs S^\dag_r \bar {\cal I}^r\Big) 
+ \mbox{h.c.} \Bigg\} +M_4^2\Big(\pp{a} \pb{a^\ast}  \Big)\times  
\nn\\
& \Big(\lb m_{3/2}\lb^2 {\cal S}^{a^\ast}{}_a +\frac {1}{m_p^2}\eKs
 \big[ \bar m_{3/2} S^p \; ({\cal S}_p)^{a^\ast}{}_a + {\mbox h.c.}\big]+\frac 1 {m_p^4} \eK
   S^p S^\dag_q \; ({\cal S}^q{}_p)^{a^\ast}{}_a \Big)\nn\\
 &+ \frac1 {m_p^2}\eK \Big(M_4^2
 \pp{a} \pb{a^\ast}  \lag \Lambda^{a^\ast}{}_a \rag +\bS{p} \bSb{p} \Big)\times
\nn\\
&  \Big(\sum \limits_ r \lb {\cal I}_r\lb^2+ M_4^3\bar{{\cal I}}^r \partial_r \omega_0+M_4^3 {\cal I}_r \partial^r \bar \omega^0\Big) +\Big( \langle S^pS_p^\dag\rangle + \langle M_4^2\varphi^\dag_{a^\ast}\Lambda^{a^\ast}{}_b\varphi^b\rangle  \Big)\times
\nn\\ 
&  \Big( 3\lb m_{3/2}\lb^2 - \lb m'_{3/2}\lb^2 -\frac{1}{2m_p^2}\eKs \big( \bar{m}'_{3/2}S^q\mathfrak{d}^z {\cal I}_q + \bar{m}_{3/2}{\cal I}_q\big(\langle S^q\rangle-2S^q\big) + \mbox{h.c.}\big)\Big)  \nn\\
&+ \eKs\Bigg\{\bar m_{3/2}  M_ 4^3 R^b{}_a \pp{a} \partial_b \omega_0 +
 \frac{M_4^3} {m_p^2} \eKs (R^p)^b{}_a\pp{a} S^\dag_p\partial_b \omega_0 
 \nn\\
 & \hskip 0.5truecm + \big[\bar m_{3/2} + \frac 1{m_p^2} \eKs S^\dag_q \bar {\cal I}^q\big] \bS{p}
\big[{\cal I}_p +M_4^3 \partial_p \omega_0\big] +
\frac{M_4^3}{m_p^2}\eKs \bSb{p} \bar {\cal I}^p \Delta \omega_0 + \mbox{h.c.}
 \Bigg\} \nn\\
&+ \eKs M_4 ^3\Big(\Delta \mathfrak{d}_z \omega_0\big[
\frac {\bar m_{3/2}}{\bar\xi_{3/2}}  + \frac1 {m_p^2} \eKs S^\dag_q \mathfrak{d}^z \bar {\cal I}^q\big]
-3 \Delta  \omega_0\big[\bar m_{3/2} + \frac1 {m_p^2} \eKs S^\dag_q \bar {\cal I}^q\big]+ \mbox{h.c.} \Big)\nn \\
& + \frac{1}{2m_p^4}\lb {\cal I}_p\lb^2\eK\Big(M_4^2
 \pp{a} \pb{a^\ast}  \lag \Lambda^{a^\ast}{}_a \rag +\bS{p} \bSb{p} \Big)^2  \nn\\
 & - M_4^2\Bigg(\Big(  m'_{3/2} + \frac1 {m_p^2} \eKs S^p \mathfrak{d}_z{\cal I}_p \Big)\langle \varphi^\dag\partial^z\Lambda\varphi\rangle\Big( \bar m_{3/2} + \frac1 {m_p^2} \eKs S^\dag_q \bar {\cal I}^q \Big) +\mbox{h.c.} \Bigg) \nn
\end{align} }
where we define $\xi_{3/2} = m_{3/2}/m'_{3/2}$ and 
{\small 
\beqa
{\cal S}^{a^\ast}{}_a&=&\frac 1 {\lb \xi_{3/2}\lb^2}\Big( \lag \partial^z \Lambda^{a^\ast}{}_b (\Lambda^{-1})^{b}{}_{b^\ast} \partial_z \Lambda^{b^\ast}{}_a -
\partial^z \partial_z \Lambda^{a^\ast}{}_a\rag\Big) + \lag \Lambda^{a^\ast}{}_a\rag\big(\frac 1{\lb \xi_{3/2}\lb^2} -2\big)\nn\\
({\cal S}_p)^{a^\ast}{}_a&=&\frac 1 {\bar \xi_{3/2}}\Big( \lag \partial^z \Lambda^{a^\ast}{}_b (\Lambda^{-1})^{b}{}_{b^\ast} \partial_z \Lambda^{b^\ast}{}_a -
\partial^z \partial_z \Lambda^{a^\ast}{}_a\rag\Big)
\mathfrak{d}_z {\cal I}_p +
\lag \Lambda^{a^\ast}{}_a\rag\Big(\frac 1 {\bar \xi_{3/2}}
 \mathfrak{d}_z {\cal I}_p  -2 {\cal I}_p \Big)\nn\\
( {\cal S}^q{}_p)^{a^\ast}{}_a&=&\Big( \lag \partial^z \Lambda^{a^\ast}{}_b (\Lambda^{-1})^{b}{}_{b^\ast} \partial_z \Lambda^{b^\ast}{}_a-
\partial^z \partial_z \Lambda^{a^\ast}{}_a\rag\Big)
\mathfrak{d}_z{\cal I}_p\mathfrak{d}^z \bar {\cal I}^q+
\lag \Lambda^{a^\ast}{}_a\rag (\mathfrak{d}_z{\cal I}_p\mathfrak{d}^z \bar {\cal I}^q-2 {\cal I}_p\bar {\cal I}^q\big)\ .\nn\eeqa}
\end{appendix}
and
{\small 
\begin{gather}
T=\frac 1 {\lb \xi_{3/2}\lb^2} -2\ , \quad T_p= \frac 1{\bar \xi_{3/2}} \mathfrak{d}_z {\cal I}_p -2 {\cal I}_p\ ,\quad T^p{}_q= \mathfrak{d}_z {\cal I}_q  \mathfrak{d}^z \bar {\cal I}^q-
2 {\cal I}_q  \bar {\cal I}^q \ , \\
R^a{}_b = \delta^a_b - \frac 1{\bar \xi_{3/2}} \lag (\Lambda^{-1})^a{}_{b^\ast} \partial_z \Lambda^{b^\ast}{}_b\rag\ ,\quad (R^p)^a{}_b= \bar {\cal I}^p \delta^a{}_b -  \mathfrak{d}^z \bar{\cal I}^p \lag (\Lambda^{-1})^a{}_{b^\ast} \partial^z \Lambda^{b^\ast}{}_b\rag\ .\nn \end{gather}}
\bibliography{bib}
%\nolinenumbers
\end{document}